\documentclass{aa}  
\usepackage{graphicx}
\usepackage{color}
\usepackage{graphicx}
\usepackage{lscape}
\usepackage{natbib}
\usepackage[varg]{txfonts}
\usepackage{amssymb}
\usepackage{stmaryrd}
\usepackage{url}
\usepackage{xspace}
\usepackage{color}
\usepackage[usenames,dvipsnames]{xcolor}
\bibpunct{(}{)}{;}{a}{}{,}
\usepackage{longtable}
\usepackage{lscape}
\newcounter{Rco}
\newcommand{\Ionst}[1]{\setcounter{Rco}{#1}\Roman{Rco}}
\newcommand{\Ion}[2]{\mbox{#1\,{\scriptsize\Ionst{#2}}}}
\newcommand{\Ionw}[3]{\mbox{#1\,{\scriptsize\Ionst{#2}}~$\lambda\,#3$\,\AA}}

\newcommand{\Ionww}[3]{\mbox{#1\,{\scriptsize\Ionst{#2}}~$\lambda\lambda\,#3$\,\AA}}

\newcommand{\logg}{\mbox{$\log g$}\xspace}
\newcommand{\loggw}[1]{\mbox{$\log g\hspace{-0.5mm} =\hspace{-0.5mm}  #1$}}

\newcommand{\se}[1]{\mbox{Sect.\,\ref{#1}}}

\newcommand{\Teff}{\mbox{$T_\mathrm{eff}$}\xspace}

\newcommand{\ebv}{$E_\mathrm{B-V}$\xspace}

\newcommand{\Lsol}{$L_\odot$}
\newcommand{\Msol}{$M_\odot$}

\newcommand{\draft}[1]{
}
\draft{Draft Version: \today.}
\usepackage{amstext}

\begin{document}

\title{RV variable, hot post-AGB stars from the MUCHFUSS project}
\subtitle{- Classification, atmospheric parameters, formation scenarios}

\author{N.~Reindl \inst{1}
   \and S.~Geier \inst{2, 3, 4}
   \and T.~Kupfer \inst{5}
   \and S.~Bloemen \inst{6}
   \and \\ V.~Schaffenroth \inst{7,3}
   \and U.~Heber \inst{3}
   \and B.~N.~Barlow \inst{8}
   \and R.~H.~\O stensen \inst{9}}

\institute{Institute for Astronomy and Astrophysics, Kepler Center for Astro and Particle Physics, Eberhard Karls University, Sand 1, 72076, T\"ubingen, Germany\\ \email{reindl@astro.uni-tuebingen.de}
\and Department of Physics, University of Warwick, Coventry CV4 7AL, UK
\and Dr.~Karl~Remeis-Observatory \& ECAP, Astronomical Institute, Friedrich-Alexander University Erlangen-Nuremberg, Sternwartstr.~7, D 96049 Bamberg, Germany
\and ESO, Karl-Schwarzschild-Str. 2, 85748 Garching bei M\"unchen, Germany
\and Division of Physics, Mathematics, and Astronomy, California Institute of Technology, Pasadena, CA 91125, USA
\and Department of Astrophysics/IMAPP, Radboud University Nijmegen, P.O. Box 9010, 6500 GL Nijmegen, The Netherlands
\and Institute for Astro- and Particle Physics, University of Innsbruck, Technikerstr. 25/8, 6020 Innsbruck, Austria
\and Department of Physics, High Point University, 833 Montlieu Avenue, High Point, NC 27262, USA
\and Institute of Astronomy, KU Leuven, Celestijnenlaan 200D, B-3001 Heverlee, Belgium}

\date{Received -- \ Accepted --}

\abstract{In the course of the MUCHFUSS project we have recently discovered four radial velocity (RV) variable, 
hot (\Teff\ $\approx 80\,000 - 110\,000$\,K) post-asymptotic giant branch (AGB) stars. Among them, 
we found the first known RV variable O(He) star, the only second known RV variable PG\,1159 close binary 
candidate, as well as the first two naked (i.e., without planetary nebula (PN)) H-rich post-AGB stars of 
spectral type O(H) that show significant RV variations. 
We present a non-LTE spectral analysis of these stars along with one further O(H)-type star whose 
RV variations were found to be not significant. We also report the discovery 
of an far-infrared excess in the case of the PG\,1159 star. 
None of the stars in our sample displays nebular emission lines, which can be explained well in 
terms of a very late thermal pulse evolution in the case of the PG\,1159 star.
The ``missing'' PNe around the O(H)-type stars seem strange, since we find that several 
central stars of PNe have much longer post-AGB times. Besides the non-ejection of a PN, the 
occurrence of a late thermal pulse, or the re-accretion of the PN in the 
previous post-AGB evolution offer possible explanations for those stars not harbouring a PN (anymore).
In case of the O(He) star J0757 we speculate that it might have been previously part of a compact He 
transferring binary system. In this scenario, the mass transfer must have stopped after a certain time, 
leaving behind a low mass close companion that could be responsible for the extreme RV shift of 
$107.0 \pm 22.0$\,km/s measured within only 31\,min.}


\authorrunning{Reindl et al.}
\titlerunning{RV variable, hot post-AGB stars}

\maketitle

\section{Introduction}
\label{sect:introduction} 

The discovery and analysis of close binaries throughout all evolutionary stages plays an important 
role in various astronomical fields. Due to the radiation of gravitational waves, very close (periods 
less than a few hours) binary systems may merge within a Hubble time. If the system 
contains two sufficiently massive white dwarfs or a subdwarf star and a sufficiently massive white 
dwarf, this evolutionary path can lead to a type Ia supernova \citep{Webbink1984}, the so-called double 
degenerate scenario. 
If the total mass of the two merging stars does not exceed the Chandrasekhar limit, He-dominated 
stars, i.e., R Coronae Borealis stars (RCB), extreme helium (EHe) stars, He-rich subdwarf O (He-sdO) 
stars or O(He) stars may be produced 
\citep{Webbink1984, Iben1984, justham2011, zhangetal2012a, zhangetal2012b, Zhang2014, Reindletal2014b}. 
The existence of planetary nebulae (PNe) around every other O(He) star and the [WN] type central stars, however, 
questions the possibility that all He-dominated stars have a merger origin. Those stars might have 
lost their H-rich envelope via enhanced mass-loss during their post-asymptotic giant branch (AGB) 
evolution, possibly triggered by a close companion \citep{Reindletal2014b}.\\
Close binary evolution is also needed to explain the formation of low mass He-core white dwarfs 
($M < 0.5$\,\Msol), because the evolutionary time-scale of the supposed low-mass single main-sequence 
progenitor stars significantly exceeds the Hubble time. The orbital energy released during a common 
envelope phase, is supposed to rapidly expel the envelope of the pre-low mass white dwarf and may 
terminate the growth of the He-core before it reaches a sufficient mass for He ignition 
\citep{Paczynski1976, Webbink1984, IbenTutukov1986}.\\
Also the formation of hot subdwarfs (sdO/B), located at the (post-) extreme horizontal branch, may best be 
understood in terms of close binary evolution. The progenitors of these stars are expected to have 
undergone a strong mass-loss already on the red giant branch (RGB), which was likely triggered by the 
close companion. If the mass-loss was sufficiently high, these stars leave the RGB and ignite He only 
while evolving to or descending the white dwarf cooling curve, and will consequently, reach a hotter 
zero-age horizontal position \citep{Brown2001, lanzetal2004}.\\
Finally, close binaries are proposed to play a crucial role in the formation of asymmetrical PNe, 
which make up the great majority (80--85\%, \citealt{Parkeretal2006}) of all PNe. The so-called binary 
hypothesis \citep{DeMarco2009} states that a companion is needed to account for the non-spherical 
shapes of these PNe. Several research groups are trying to determine the frequency of binary central stars 
of PNe (CSPNe) and the properties of these systems. Most of the known close binary CSPNe were detected 
via periodic flux variability due to eclipses, irradiation, and/or ellipsoidal deformation 
\citep{Bond2000, Miszalski2009}. This technique is, however, biased against binaries with periods longer 
than about two weeks, against binaries with low inclination angles, and against companions 
with small radii \citep{DeMarco2008}. Large radial velocity (RV) surveys of CSPNe are telescope-time 
expensive and not always straightforward, because most of the strongest photospheric lines are blended with 
nebular emission lines and the extraction of the pure stellar spectrum may often not be successful in the 
case of very compact nebulae.\\
Fortunately, several hundreds of hot subluminous stars, which generally do not display nebular lines, were 
observed in the course of the Sloan Digital Sky Survey (SDSS). Because SDSS spectra are co-added from at least 
three individual integrations (with a typical exposure time of 15 min), they offer the ideal 
basis for the search for RV variations. The MUCHFUSS (Massive Unseen Companions to Hot Faint Underluminous 
Stars from SDSS) survey made use of that to search for hot subdwarf stars with massive compact companions 
\citep{Geieretal2011a}.  For that purpose, hot subdwarfs were selected from the SDSS DR7 by color (i.e., 
they were required to be point sources with $u -g < 0.4$ and $g - r < 0.1$) and visual inspection of their 
spectra in order to distinguish them from hot white dwarfs or extragalactic sources. Those stars, whose SDSS 
sub-spectra revealed high RV variations, were selected as candidates for follow-up spectroscopy to derive the 
RV curves and the binary mass functions of the systems. In this way, more than 100 RV-variable subdwarfs have 
been found \citep{Geieretal2015}.\\
Hot post-AGB stars directly overlap with hot subdwarfs of O-type (sdO) in the color-color plane. 
If hot post-AGB stars are not surrounded by a compact nebula and/or do not suffer from still ongoing strong 
mass loss (i.e., weak emission line stars, Wolf Rayet type central stars, whose spectra show or are even 
dominated by emission lines), their spectra may look indistinguishable from the less luminous sdO stars.  
About three-quarter of all hot post-AGB stars display a normal H-rich composition and in case their spectra 
are dominated by absorption lines, they can be classifed as O(H) or hgO(H) stars (the latter case applies 
to high gravity stars which are positioned on the white dwarf cooling track, \citealt{Mendez1991}). 
Hot, H-deficient stars, which show predominatly absorption lines and lie in the post-AGB region, are either 
classified as O(He) stars (typically showing more than 90\% He, by mass), or PG\,1159 stars, whose spectra 
show besides He also strong lines of carbon. 
By eye, the spectra of O(H), O(He), and PG\,1159 stars may look very similar to those of the less luminous 
sdO, He-sdO, and C-strong lined He-sdO stars, respectively. Thus, the sample of \cite{Geieretal2015} also contains 
a small fraction of hot post-AGB candidates, namely \object{SDSS\,J075732.18+184329.3} 
(here after J0757), \object{SDSS\,J093521.39+482432.4} (J0935), \object{SDSS\,J100019.98-003413.3} (J1000), 
\object{SDSS\,J155610.40+254640.3} (J1556), and \object{SDSS\,J204623.12-065926.8} (J2046). Previous analyses 
of these stars indicated very high effective temperatures \citep{Werneretal2014, Geieretal2015}. Because of 
the rareness of such objects as well as their importance for our understanding of the late, hot stages of 
stellar evolution we decided to subject these five stars to a more comprehensive analysis and discussion of 
their evolutionary status.\\
The paper is organized as follows. First, we describe the observations and the discovery of RV variations 
(\se{sect:rvv}). Then, we determine the parameters of the stars in \se{sect:parameter}. In \se{sect:irexcess} follows 
a search for a PN and an infrared excess. The evolutionary status of the stars is discussed in \se{sect:discussion} 
and finally we conclude in \se{sect:conclusions}.

\begin{figure}[ht]
\resizebox{\hsize}{!}{\includegraphics{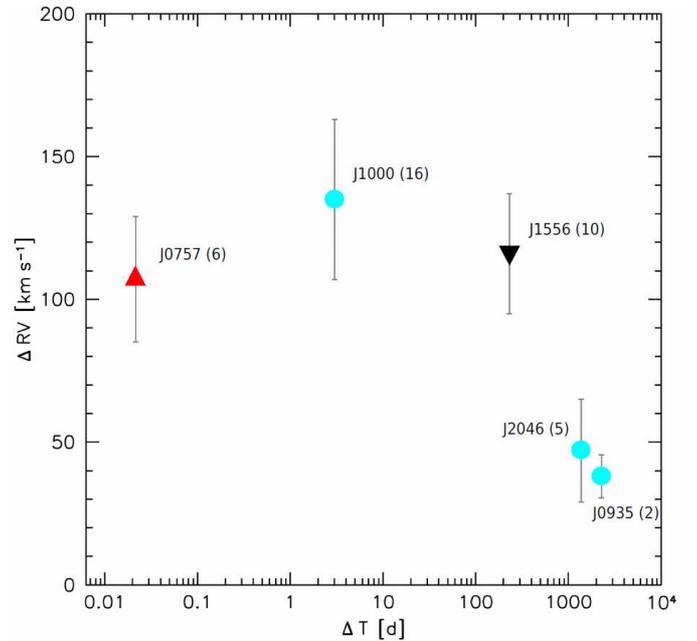}}
\caption{Highest RV shift between individual spectra plotted against time
difference between the corresponding observing epochs for the O(He)
star J0757 (red, triangle), the PG1159 star J1556 (black, inverted triangle), and the three
O(H) stars J1000, J2046, and J0935 (blue, filled circles). The numbers in the
brackets correspond to the number of epochs each star was observed. Note that 
the RV shifts of J2046 were found to be not significant.}
\label{dTdRV_postagb}
\end{figure}

\section{Observations and detection of radial velocity variations}
\label{sect:rvv} 

Individual observations and RV measurements of the five stars are listed in \cite{Geieretal2015}. 
The RVs were measured by fitting a set of mathematical functions (Gaussians, Lorentzians, and 
polynomials) to the spectral lines using the FITSB2 routine \citep{Napiwotzkietal2004}. The fraction 
of false detections produced by random fluctuations and the significance of the measured RV variations 
were calculated as described in \cite{Maxted2001}.\\
In the spectra of J0757, J0935, J1000, and J1556 we discovered significant RV shifts, while the RV 
variations in the five SDSS sub-spectra of J2046 were found to be most likely caused by random 
fluctuations (false-detection probability p is larger than 0.01\%). In Fig.~\ref{dTdRV_postagb} we 
show their highest RV shift between individual spectra plotted against time difference between the 
corresponding observing epochs. 
Most impressive, we discovered a maximum RV shift of $107.0 \pm 22.0$\,km/s within only 31\,min 
in the six SDSS sub-spectra (taken in two consecutive nights) of J0757. J1000 was observed 
16 times in the course of the SDSS and reveals a maximum RV shift of $135.0 \pm 28.0$\,km/s. 
In addition to the nine SDSS observations of J1556, we have obtained one medium resolution spectrum with 
TWIN at the Calar Alto  3.5\,m telescope and found a maximum RV shift of $116.0 \pm 21.0$\,km/s.
For J0935 second epoch spectroscopy was obtained with WHT/ISIS. The maximum RV shift is 
$38.0 \pm 7.5$\,km/s.\\
To verify the close binary nature of the stars, we tried to derive orbital periods and RV semiamplitudes 
of the objects with more than ten epochs as described in \cite{Geieretal2015}. 
Unfortunately, in both cases the RV datasets were insufficient to put any meaningful constraints on the 
orbital parameters.

\section{Stellar parameters and distances}
\label{sect:parameter} 

In this section we first perform a non-LTE spectral analysis to derive the effective 
temperatures, surface gravities and the chemical composition of the five stars (\se{subsect:specana}). 
These parameters are then used in \se{subsect:masses} to derive their masses, luminosities, and 
distances. 

\subsection{Spectral analysis}
\label{subsect:specana} 

Quantitative spectral analysis is the key to derive the surface parameters of the stars and to 
understand their evolutionary status. Our spectral analysis was based on the SDSS observations 
of the stars. The single spectra have been corrected for their orbital motion and coadded.
For the model calculations we employed the T{\"u}bingen non-LTE 
model-atmosphere package (TMAP\footnote{http://astro.uni-tuebingen.de/~TMAP}, 
\citealt{werneretal2003, rauchdeetjen2003}) which allows to compute plane-parallel, non-LTE, fully 
metal-line blanketed model atmospheres in radiative and hydrostatic equilibrium. Model atoms were 
taken from the T{\"u}bingen model atom database TMAD\footnote{http://astro.uni-tuebingen.de/~TMAD}.
To calculate synthetic line profiles, we used Stark line-broadening tables provided by \cite{Barnard1969} 
for \Ionww{He}{1}{4026, 4388, 4471, 4921}, \cite{Barnard1974} for \Ionw{He}{1}{4471}, and \cite{Griem1974} 
for all other \ion{He}{I} lines. For \ion{He}{II} we used the tables provided by \cite{Schoening1989}, for 
H\,{\sc i} tables provided by \cite{TremblayBergeron2009}, and for \Ion{C}{4} tables provided by 
\cite{Dimitrijevic1991} and \cite{Dimitrijevic1992}. 
\\

\subsubsection{H-rich stars}
\label{subsect:Hspecana}
 
The spectra of J0935, J1000, and J2046 are dominated by hydrogen Balmer lines. Spectal lines of He\,{\sc ii} 
are of moderate strength, but no He\,{\sc i} lines can be detected. However, the signal to noise of the 
observations is too poor (S/N $\approx$ 20-30) to exclude their presence from the outset. 
To derive effective temperatures, surface gravities, and He abundances of these stars, we calculated a model 
grid spanning from $T_{\rm eff}=40\,000-140\,000\,{\rm K}$ (step size 1250\,K) and $\log{g}=4.75-6.5$ (step size 0.25\,dex) for 
six different He abundances (He$=0.1, 0.2, 0.3, 0.4, 0.5$, and $0.6$, by mass). Models above the Eddington 
limit (i.e, $T_{\rm eff}>90\,000$\,K for \loggw{4.75}, $T_{\rm eff}>100\,000$\,K for \loggw{5.00}, and 
$T_{\rm eff}>120\,000$\,K for \loggw{5.25)} were not calculated. For He\,{\sc ii} 20 levels were considered in 
non-LTE, and for He\,{\sc i} 29 levels if $T_{\rm eff}<100\,000$\,K, and 5 levels if $T_{\rm eff} \ge 100\,000$\,K. 
For H\,{\sc i} 15 levels were considered in non-LTE.
\begin{figure}[ht]
 \centering
\resizebox{\hsize}{!}{\includegraphics{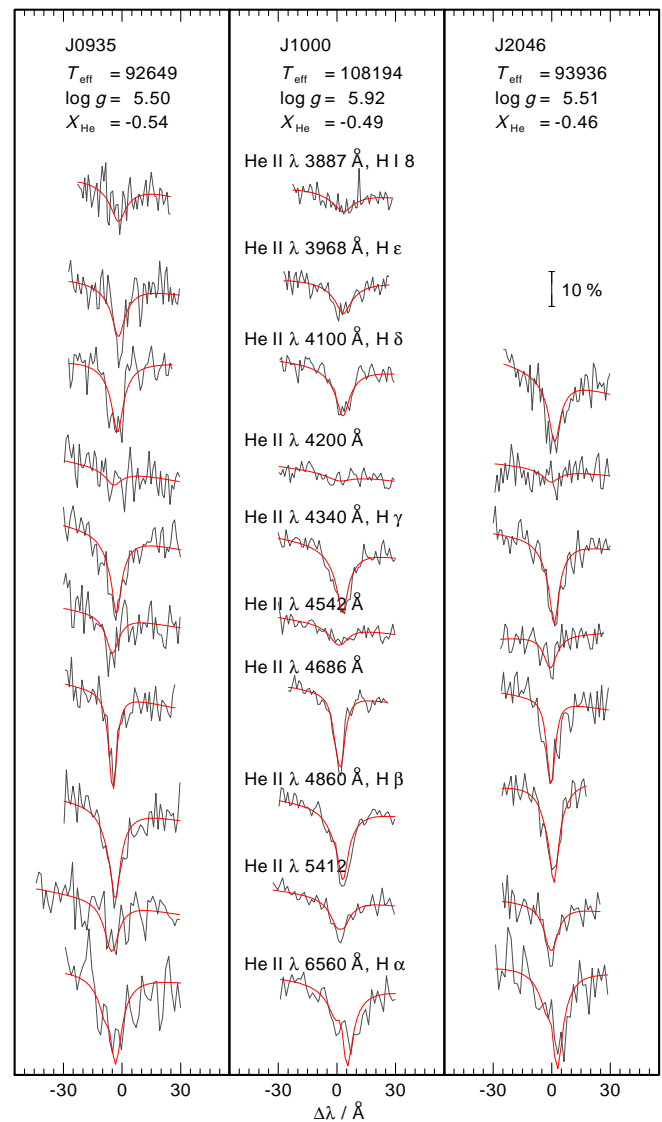}}
\caption{Balmer and He\,{\sc ii} lines used to derive \Teff, \logg, and the He abundance (given in logarithmic mass fractions) 
of the three O(H)-type stars. The observations are shown in grey, the best fit models in black (red in online version). The 
vertical bar indicates 10\,\% of the continuum flux.}
\label{fig:OH}
\end{figure}
The parameter fit was then performed by means of a $\chi^2$ minimization technique with SPAS (Spectrum 
Plotting and Analysing Suite, \citealt{Hirsch2009}), which is based on the FITSB2 routine \citep{Napiwotzki1999}. 
We fitted all Balmer and He\,{\sc ii} lines detected in the SDSS spectra of these stars. 
\Ionw{He}{2}{3968}\,/\,H\,$\epsilon$ and \Ionw{He}{2}{3887}\,/\,\Ionw{H}{1}{3889} were not fitted in case of 
J2046 due to the poor quality of the blue part of the spectra. Our best fits are shown Fig.~\ref{fig:OH} and the 
results of our analysis are summarized in Table~\ref{tab:parameters}.\\
The spectra were previously analysed by \cite{Geieretal2015} based on a model grid calculated by 
\cite{stroeeretal2007} and we find that they generally underestimated \Teff\ by 5 to 15\%. In addition, our 
results for J1000 lie outside the grid used by \cite{Geieretal2015}. While the effects of the smaller step 
size in \Teff\ ($1250$\,K instead of $5\,000$\,K) of our grid, are relatively small (\Teff is usually 
underestimated by 1\% when using a finer grid), the main influence can be attributed in the larger extent 
towards higher \Teff\ ($140\,000$\,K instead of $100\,000$\,K), as well as in the use of the Stark 
line-broadening tables provided by \cite{TremblayBergeron2009} instead of those provided by 
\cite{Lemke1997}.

\begin{figure}[ht]
  \resizebox{\hsize}{!}{\includegraphics{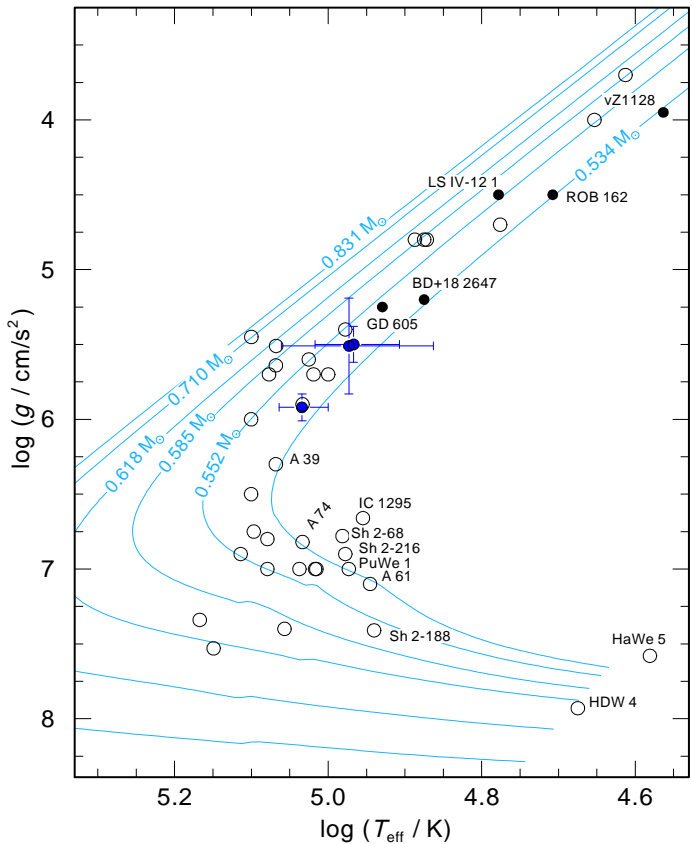}}
\caption{The locations of our three O(H) stars (filled, blue circles with error bars) are compared with H-burning 
post-AGB evolutionary tracks (blue lines, labeled with stellar masses, \citealt{MillerBertolami2015}). Also shown 
are the locations of naked O(H)-type stars (filled circles, 
\citealt{Chayeretal2015, Fontaine2008, BauerHusfeld1995, HeberHunger1987, Heber1986}), as well as the locations of 
H-rich CSPNe (open circles, \citealt{ZieglerPhD2012, heraldbianchi2011, Gianninas2010, Rauch2007, Napiwotzki1999}).}
\label{fig:tefflogg_postagb_H}
\end{figure} 

It is important to note, that the errors given in Table~\ref{tab:parameters} correspond to formal fitting errors 
only. The neglect of metal opacities in the model-atmosphere calculations can lead to significant systematic 
errors in the inferred atmospheric parameters of very hot stars. \cite{Latour2010, Latour2015} showed that for 
sdO stars hotter than about $70\,000$\,K, \Teff and \logg are typically underestimated by 5 to 10\% and 0.15\,dex, 
respectively.\\ 
However, even considering these effects, all stars lie clearly in the post-AGB region of the $\log$\Teff\ $-$ \logg\ 
diagram, shortly before they will reach their maximum \Teff, to then cool down as white dwarfs 
(Fig.~\ref{fig:tefflogg_postagb_H}). Their He abundances are slightly supersolar ($1.2 - 1.4\,\times$ solar, solar 
abundance according to \citealt{asplundetal2009}). In order to distinguish these stars from sdO stars, 
which are usually associated with AGB-manqu\'{e} stars, which are stars that fail to evolve through the AGB,  
we employ the spectral classification scheme of \cite{Mendez1991} and classify them as O(H) type stars.

\subsubsection{H-deficient stars}
\label{subsect:Hdefspecana}

\begin{figure*}[ht]
 \centering
  \includegraphics[width=\textwidth]{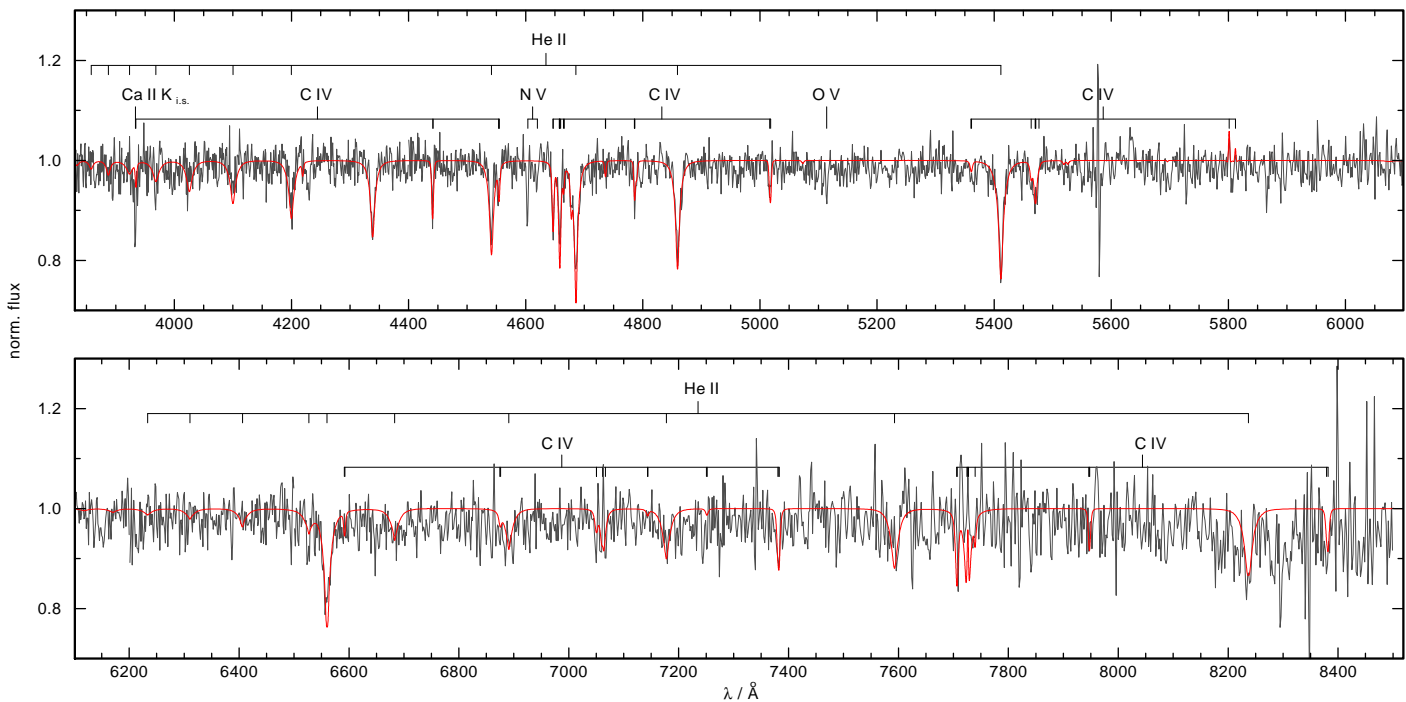}
\caption{Coadded and normalized SDSS spectrum (grey) of the PG\,1159 star SDSS\,J155610.40+254640.3. The best fit model 
(black, red in online version) including He and C lines is superimposed and the positions of identified spectral lines 
are indicated.}
\label{J1556}
\end{figure*}

J0757 has been analysed by \cite{Werneretal2014}, who derived $T_{\rm eff}=80\,000$\,K, $\log{g}=5.00$ 
and a C abundance of $C = 0.006$ (by mass). Thus, the star belongs to the class of C-rich O(He) 
stars.\\ 
The C lines in the spectrum of J1556 are stronger and more numerous compared to the ones found in the 
spectrum of J0757. In Fig.~\ref{J1556} the co-added SDSS spectrum of J1556 is shown. Besides photospheric 
lines of He\,{\sc ii} and C\,{\sc iv}, we could identify lines of N\,{\sc v} and O\,{\sc v}. Constraints 
on the effective temperature can already be put from the absence of He\,{\sc i} lines 
($T_{\rm eff} \ge 70\,000\,{\rm K}$) and the \Ionww{C}{4}{5801, 5812} lines, which appear in emission 
($T_{\rm eff} \ge90\,000\,{\rm K}$). In addition, the fact that the N\,{\sc v} lines at $4604\,{\rm \AA}$ 
and $4620\,{\rm \AA}$ appear in absorption allows us to put an upper limit of 
$T_{\rm eff} \le 115\,000\,{\rm K}$ \citep{rauchetal1994}.\\
For the spectral analysis of J1556 we firstly calculated a model grid including only the opacities of H, He, and C.
For H and He non-LTE levels were used as described in \se{subsect:Hspecana}. The C model atom includes the 
ionization stages {\sc iii}-{\sc v} and 30, 54, and 1 non-LTE levels were considered for \Ion{C}{3}, 
\Ion{C}{4}, and \Ion{C}{5}, respectively. Our model grid spans $T_{\rm eff}=90\,000-120\,000\,{\rm K}$ and 
$\log{g}=5.0-6.7$ and C abundances in the range of $0.33-0.01$ (by mass) were considered, The grid is, 
however, not complete due to the occurrence of numerical instabilities. The fitting of the spectrum was then 
carried out in an iterative process, in which the parameters of the model have been changed successively 
until a good agreement with the observation was achieved. The best fit was found for 
$T_{\rm eff}=100\,000\,{\rm K}$, $\log{g}=5.33$, and $C=0.15$. No traces of H could be detected. 
Unfortunately, we did not succeed in calculating numerically stable models that also included 
the opacities of N and O at such low surface gravities, which precluded the abundances determination of 
those elements.\\
The C abundance lies well above the limit set by \cite{Werneretal2014} to distinguish between the PG\,1159 
stars and the He-dominated DO white dwarfs and O(He) stars, which only show up to $C = 0.03$. Therefore, we 
classify J155610.40+254640.3 as a PG\,1159 star. It is worthwhile to note, that this is also the PG\,1159 star 
with the lowest surface gravity known so far. Interestingly, its C abundance is rather low compared to the 
majority of the luminous PG\,1159 stars (that is they have \logg\ $< 7.0$) which typically show $C = 0.5$ 
(see Fig.\,4 in \citealt{Reindletal2014c} for comparison).

\begin{table*}
\caption{Stellar Parameter. He abundances are given in logarithmic mass fractions.}
\begin{tabular}{llllllllll}
\hline\hline
\noalign{\smallskip}
name                       & Spectral   & $T_{\rm eff}$ & $\log{g}$        & $X_{\mathrm{He}}$         & $M$                 & $\log (L/$\Lsol) &  $d$     & $z$       \\
                           & type       & [K]           &                 &                         & [\Msol]             &                      &  [kpc]  & [kpc]    \\
\noalign{\smallskip}
\hline
\noalign{\smallskip}                                                                                                                                                                                                      
J0935 & O(H)         & \,\,\,$92649 \pm 11358$           &  $5.50 \pm 0.12$ & $-0.46 \pm 0.09$ & $ 0.54^{+0.05}_{-0.01}$          & $3.5^{+0.4}_{-0.2}$ &  $16.7\,^{+2.7}_{-2.2}$      & $\,12.1\,^{+1.8}_{-1.6} $ \\
\noalign{\smallskip}                                                                                                                                                                                                      
J1000 & O(H)         & $108194 \pm \,\,\,8356$           &  $5.92 \pm 0.09$ & $-0.49 \pm 0.06$ & $ 0.54^{+0.02}_{-0.01} $         & $3.4^{+0.2}_{-0.1}$ &  $\,\,\,8.0\,^{+0.9}_{-0.8}$ & $\,\,\,\,5.2\,^{+0.6}_{-0.5} $ \\
\noalign{\smallskip}                                                                                                                                                                                                      
J2046 & O(H)         & \,\,\,$93936 \pm 21166$           &  $5.51 \pm 0.32$ & $-0.54 \pm 0.16$ & $ 0.54^{+0.08}_{-0.01} $         & $3.5^{+0.5}_{-0.2}$ &  $10.8\,^{+4.8}_{-3.3} $     & $-5.2\,^{+1.6}_{-2.3} $\\
\noalign{\smallskip}
J1556 & PG\,1159     & $100000\,_{\,\,-10000}^{\,\,+15000}$ &  $5.30 \pm 0.30$ & $-0.07 \pm 0.03$ & $ 0.57^{+0.13}_{-0.05}$          & $3.8^{+0.5}_{-0.4}$ &                 $16.3\,^{+6.7}_{-4.8}$     & $\,12.3\,^{+5.1}_{-3.6} $ \\
\noalign{\smallskip}                                                                                                                                                                                                       
J0757$^{\mathrm{(a)}}$ & O(He)        & \,\,\,$80000 \pm 20000$           &  $5.00 \pm 0.30$ & $-0.01 \pm 0.01$ & $ 0.53^{+0.21}_{-0.05}$\,$^{\mathrm{(a)}}$   & $3.7^{+0.6}_{-0.3}$\,$^{\mathrm{(b)}}$    & $26.8\,^{+11.1}_{-7.8}$\,$^{\mathrm{(b)}}$   & $\,10.3\,^{+4.3}_{-3.0}$\,$^{\mathrm{(b)}} $ \\
\noalign{\smallskip}                                                                                                                                                                                                      
            &              &                                   &                  &                  & $ 0.73^{+0.27}_{-0.08}$\,$^{\mathrm{(c)}}$   & $3.9^{+1.0}_{-0.5}$\,$^{\mathrm{(c)}}$ &   $31.5\,^{+13.0}_{-9.2}$\,$^{\mathrm{(c)}}$   & $\,12.1\,^{+5.0}_{-3.5}$\,$^{\mathrm{(c)}}$ \\
\noalign{\smallskip}
\hline
\end{tabular}
\tablefoot{~\\
\tablefoottext{a}{Atmospheric parameters teken from \cite{Werneretal2014}.}
\tablefoottext{b}{Based on (V)LTP tracks of \cite{millerbertolamialthaus2007, millerbertolamialthaus2006}.}
\tablefoottext{c}{Based on double He-white dwarf merger tracks of \cite{zhangetal2012b, zhangetal2012a}.}
}
\label{tab:parameters}
\end{table*}

\subsection{Masses, luminosities, and distances}
\label{subsect:masses} 

\begin{figure}[ht]
  \resizebox{\hsize}{!}{\includegraphics{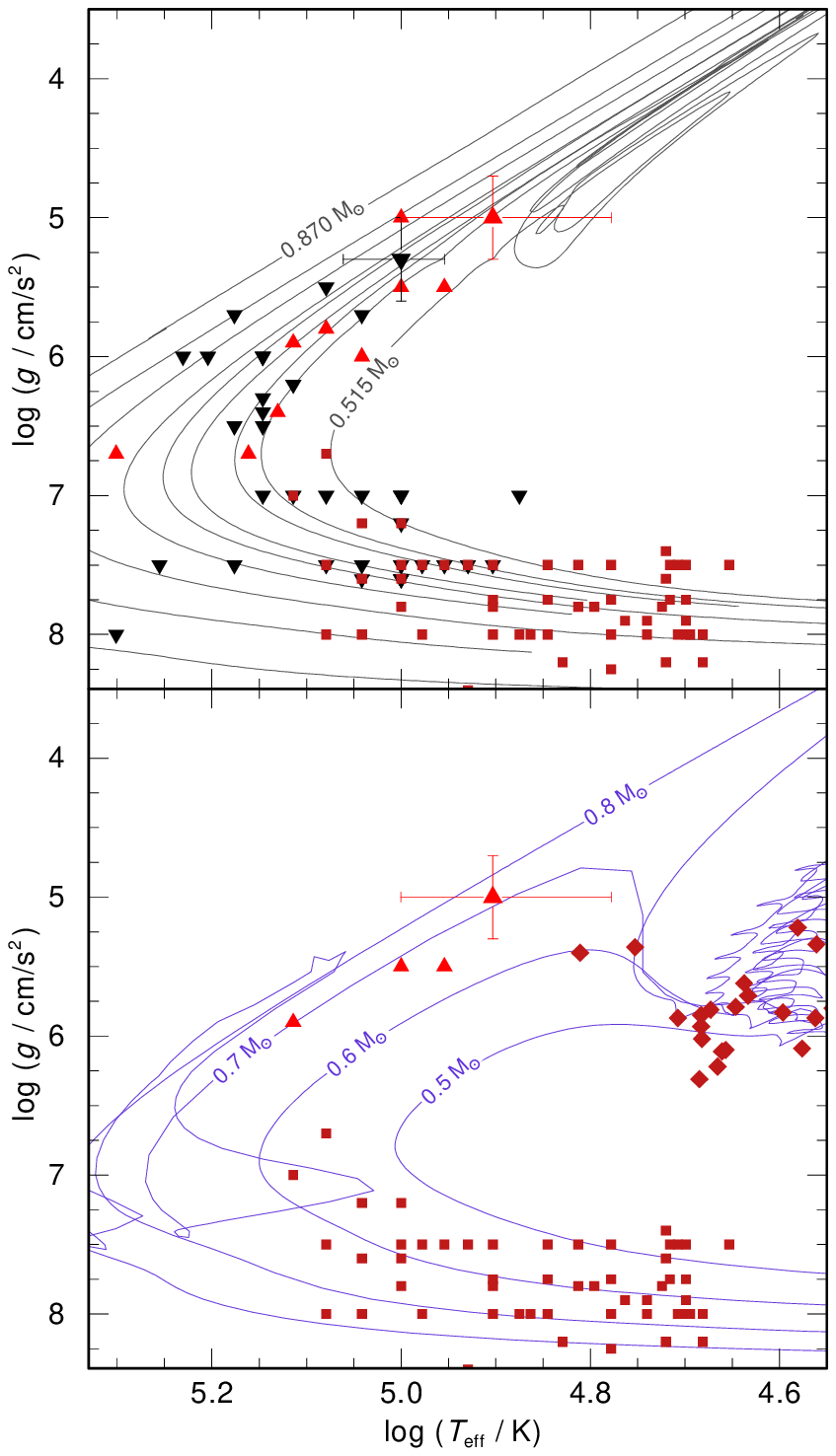}}
\caption{Locations of the O(He) star J0757 (red triangle with error bars) and the PG\,1159 star 
J1556 (black inverted triangle with error bars) and other PG\,1159 and O(He) stars (black inverted triangles 
and red triangles, respectively, \citealt{DeMarco2015, Reindletal2014b, WernerRauch2014, Werneretal2014, Gianninas2010, 
wassermannetal2010, wernerherwig2006}) in the $T_{\rm eff}-\log{g}$ diagram. Their likely successors, 
the H-deficient white dwarfs 
\citep{Reindletal2014c, Werneretal2014, mahsereci2011, huegelmeyeretal2006, dreizlerwerner1996} are 
indicated by filled, red squares. The \textit{upper panel} compares the location of these stars to 
(V)LTP evolutionary tracks (grey lines) of \cite{millerbertolamialthaus2007, millerbertolamialthaus2006}. 
For clarity, only two tracks are labeled with stellar masses. Intermediate tracks correspond to 0.530, 
0.542, 0.565, 0.584, 0.609, 0.664, 0.741\,\Msol. The \textit{lower panel} compares only the locations of 
the C-rich O(He) stars (red triangles) and C-strong lined He-sdO stars (red rhombs) from the sample of 
\cite{Geieretal2015} with post-double He white dwarf merger tracks (purple lines) of 
\cite{zhangetal2012b, zhangetal2012a}.}
\label{fig:tefflogg_postagb_Hdef}
\end{figure}

To derive the masses and luminosities of the five stars, we compared their position in the 
log \Teff\ -- \logg\ plane with different evolutionary tracks. Figure\,\ref{fig:tefflogg_postagb_H} 
shows the location of the three O(H) stars compared with H-burning post-AGB evolutionary tracks 
\citep{MillerBertolami2015}. Because these stars most likely belong 
to the Galactic halo (see below), we used tracks with $Z=0.001$. We derived $M = 0.54$\,\Msol\ for all 
O(H)-type stars, which agrees well within the error limits with the mean mass of $0.551\pm0.005$\Msol\ 
of H-rich halo white dwarfs \citep{Kalirai2012}.\\
In the upper panel of Fig.\,\ref{fig:tefflogg_postagb_Hdef} the locations of the two H-deficient stars, J0757 and 
J1556, are compared with a late thermal pulse (LTP) evolutionary track of \cite{millerbertolamialthaus2007} 
and very LTP (VLTP) post-AGB tracks of \cite{millerbertolamialthaus2006}. We derived $M = 0.57$\,\Msol\ for J1556, 
and $M = 0.53$\,\Msol\ for J0757. For J0757 we additionally derived the mass and luminosity using post-double He 
white dwarf merger evolutionary tracks of \cite{zhangetal2012b, zhangetal2012a} and found $M = 0.73$\,\Msol\ 
(lower panel of Fig.\,\ref{fig:tefflogg_postagb_Hdef}). 
Within the error limits, J0757 could also be a merger of a CO white dwarf and a He white dwarf, however, to this 
date there are no post-CO+He white dwarf merger tracks available that reach to the region of O(He) type stars.
We stress that a reliable spectroscopic mass determination for J0757 would only be possible if the evolutionary 
history of this object would be known. Therefore, the parameter listed for J0757 in Table~\ref{tab:parameters}, 
are only valid if a certain scenario is assumed.\\
Based on the derived masses, we calculated the distances of the stars by using the flux 
calibration of \citet{heberetal1984} for $\lambda_\mathrm{eff} = 5454\,\mathrm{\AA}$,
$$d\,[\mathrm{pc}]=7.11 \times 10^{4} \cdot \sqrt{H_\nu\cdot M \times 10^{0.4\, m_{\mathrm{v}_0}-\log g}},$$
\noindent
with 
$m_{\mathrm{V_o}} = m_\mathrm{V} - 2.175 c$, $c = 1.47 E_{B-V}$, and 
the Eddington flux 
$H_\nu$ at $\lambda_{\rm eff}$ 
of the best-fit model atmospheres of each star. Values for \ebv were obtained from Galactic dust extinction 
maps from \cite{Schlafly2011} and values for $m_{\mathrm{V}}$ were taken from \cite{Geieretal2015}. 
The distances range from $8.0$ to $31.5$\,kpc and deviate slightly from the values listed in \cite{Geieretal2015}. 
This is because they used different model fluxes and assumed a canonical mass of $0.6$\,\Msol\ for all 
post-AGB stars for their distance determination. We found that all five stars are located far above or below the 
Galactic plane (Table~\ref{tab:parameters}). Since the thick disk only dominates in the region 1\,kpc 
$\lesssim z \lesssim$ 4\,kpc \citep{Kordopatis2011}, we conclude that all stars from our sample probably 
belong to the Galactic halo.

\section{Search for nebular lines and infrared excess}
\label{sect:irexcess} 

All of our five stars are located in the region of CSPNe in the $\log$ \Teff\ -- \logg\ diagram, which suggests 
that they might be central stars as well. The large distances of the stars, however, imply rather small angular 
diameters of the order of one arcsec or even less (assuming a typical expansion velocity of 20\,km/s and 
distances and post-AGB times from Table~\ref{tab:ages}). Thus, it is unlikely to detect PNe around these 
stars via ground-based imaging. Therefore, it is much more promising to search for prominent nebular emission 
lines in the spectra of these stars, i.e., the [\ion{O}{III}] $\lambda\lambda$ 4959, 5007 $\AA $ emission 
lines, which are typically the strongest lines for a medium to high-excitation PN. Because of the long 
exposure times and the much better background subtraction that can be achieved in slit or fibre spectroscopy 
compared to narrow-band imaging or slitless objective prism imaging spectroscopy, SDSS spectra allow the 
detection of extremely faint PN (down to $29 - 30$\,mag\,arcsec$^{-1}$, \citealt{Yuan2013}). 
However, we could not detect any hint for nebular lines in the spectra of our five stars.

\begin{figure}[ht]
 \resizebox{\hsize}{!}{\includegraphics{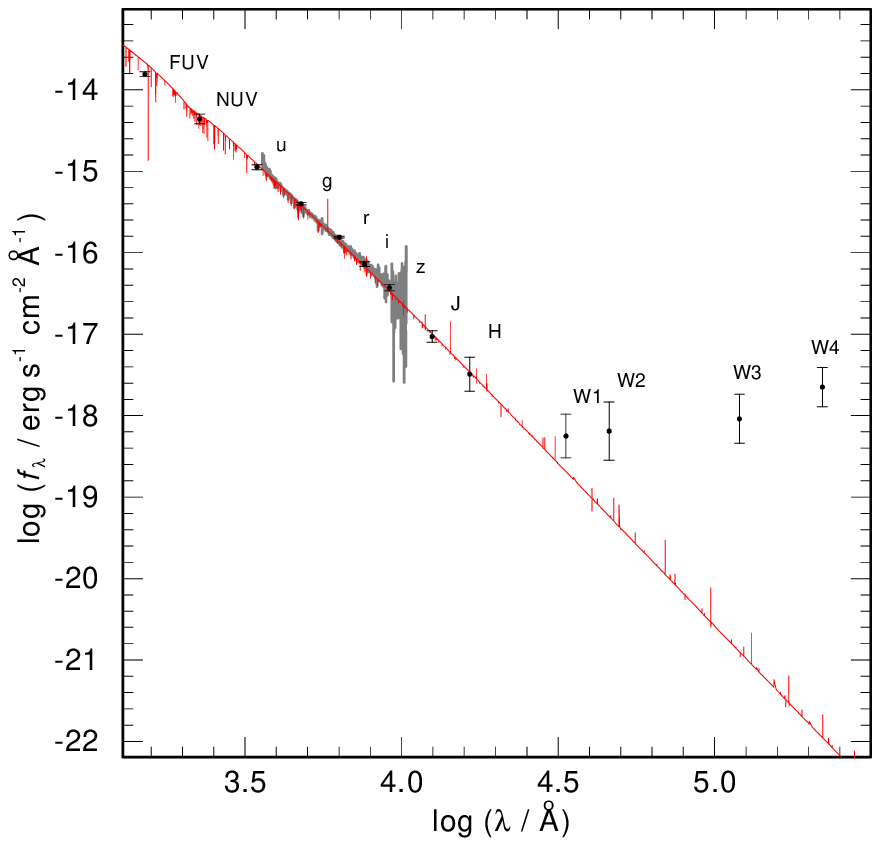}}
\caption{Colors (black dots with 3\,$\sigma$ error bars) and the SDSS spectrum (grey) of the PG\,1159 star J1556 (grey), 
compared with a reddened TMAP model SED (black, red in online version). The emission lines are of photospheric 
origin. Note that the mismatch of the reddened model flux and the GALEX $FUV$ measurement is due to missing 
metal opacities in our model atmosphere.}
\label{fig:EBV}
\end{figure}

The detection of an infrared (IR) excess can put constraints on the nature of a possible companion, or the 
presence of a circumstellar disk. To search for possible IR excess emission around our five stars, we cross matched 
their positions with the UKIDSS (UKIRT Infrared Deep Sky Survey) DR9 catalog \citep{Lawrence2013} and the WISE 
(Wide-field Infrared Survey Explorer) catalog \citep{Cutri2014}. We applied a search radius of 2 arcsec. For the 
O(H) stars J0935 and J2046 no IR colors were found, but for J1000 and the PG\,1159 star J1556 we found $J$ and $H$ 
entries. For the O(He) star J0757 we found a $J$ value. The PG\,1159 star J1556 is the only object with entries in the 
WISE catalog. In addition, GALEX $FUV$ and $NUV$ \citep{Bianchi2014} and SDSS magnitudes \citep{Ahnetal2012} were 
obtained for all objects. Magnitudes and reddening of the five stars are summarized in Table~\ref{tab:mags}.
SDSS, UKIDSS, and WISE magnitudes were converted into fluxes as outlined in 
\cite{Verbeeketal2014}. For the GALEX magnitudes 
conversions\footnote{http://asd.gsfc.nasa.gov/archive/galex/FAQ/counts\_background.html}, we used
$$f_{\mathrm{FUV}}\,[\small{\mathrm{erg}\,\,\mathrm{s^{-1}\,cm}^{-2}\mathrm{\AA}^{-1}}] = 1.40\times10^{-15}\,10^{-0.4\,\,(FUV-18.82)}$$
$$f_{\mathrm{NUV}}\,[\small{\mathrm{erg}\,\,\mathrm{s^{-1}\,cm}^{-2}\mathrm{\AA}^{-1}}] = 2.06\times10^{-16}\,10^{-0.4\,\,(NUV-20.08)}.$$
\noindent
Then, we corrected the TMAP model flux of our best fit for each star for interstellar reddening using the 
reddening law of \cite{fitzpatrick1999}. Again, we used the values for \ebv from the Galactic dust extinction maps 
of \cite{Schlafly2011}. Thereafter, we normalized the reddened model flux to the SDSS $g$ magnitude and checked if 
the other color measurements are within 3$\sigma$ in agreement with the model fluxes.\\ 
A perfect agreement with the color measurements and the reddened model flux was found for the three O(H) stars and 
the O(He) star, i.e. no evidence for an excess was found in the near IR. The PG\,1159 star J1556, however, displays 
a clear excess in all WISE colors (Fig.~\ref{fig:EBV}). Due to the lack of WISE measurements a 
search for an far-IR excess was not posible for the other four stars.
In the (W2$-$W3), (W3$-$W4) diagram, the WISE colors of J1556 agree with the bulk of PNe \citep{Kronberger2014}. 
However, as mentioned above, no nebular emission lines are visible in the SDSS spectrum (total exposure time 
$\approx 2.25$\,h).\\
An IR excess that strongly shows up in the Spitzer IRAC bands \citep{XuJura2012}, was found around some warm 
(i.e., $T_{\rm eff} \le 25\,000\,{\rm K}$) white dwarfs. This IR excess is interpreted as originating from a small 
hot disk due to the infall and destruction of single asteroids that come within the star's Roche limit \citep{Gaensicke2006}.
However, we exclude the presence of a debris disk in the case of J1556, because at such high \Teff, solids would be 
sublimated into gaseous disks \citep{vonHippel2007}. In addition the double-peaked \ion{Ca}{II} triplet, which is
the hallmark of a gaseous, rotating disk \citep{Youngetal1981, HorneMarsh1986}, cannot be detected.\\
Cold dust disks, which are located much farther from the star ($\approx 50$\,AU compared to $\le 0.01$\,AU for the 
debris disks) have been announced for a number of hot white dwarfs and CSPNe and are likely formed by mass loss from the stars 
during their AGB phase \citep{Claytonetal2014, Chuetal2011}. These disks, however, were found to be relatively cool 
($\approx 100$\,K compared $\approx 1000$\,K for the debris disks) and should show up only in the W4 band. 
The relatively flat IR excess in J1556 cannot be fitted with a single blackbody and looks very similar like in the 
case of the CSPN \object{K\,1-22}. \cite{Chuetal2011} showed that for K\,1-22, the IR excess is a superposition of the 
photospheric emission of the CSPN and a red companion as well as the dust continuum. Thus, it is likely that more than 
one source contributes to the IR excess of J1556 as well.\\
Finally, we note that the IR excess could also have an extragalactic origin. We corrected the WISE colors of J1556 for 
the stellar flux contribution and obtained $W1=18.40$, $W2=16.54$, $W3=12.23$, and $W4=8.93$. This leads to W1$-$W2 
$= 1.86$ and W2$-$W3 $=4.31$, which corresponds to the region of luminous IR galaxies or Sbc star-burst galaxies in the 
(W1$-$W2), (W2$-$W3) diagram (Fig.\,12 in \citealt{Wright2010}). The WISE-excess might therefore be caused by a chance 
alignment of such a background galaxy with J1556. However, the probability for such an alignment of two very peculiar 
classes of objects seems to be very unlikely.

\onltab{
\onecolumn
\begin{table*}
\caption{Magnitudes and reddening of the stars}
\begin{tabular}{cccccc}
\hline\hline
\noalign{\smallskip}
Object & J0935            & J1000           & J2046            & J1556  & J0757  \\
\hline
\noalign{\smallskip}
\ebv   & $\,\,\,0.018\pm0.001$ & $\,\,\,0.036\pm0.002$ & $\,\,\,0.058\pm0.003$ & $\,\,\,0.058\pm0.005$ & $\,\,\,0.030\pm0.001$\\
\noalign{\smallskip}
$FUV$ & $16.325\pm0.019$ & $15.799\pm0.019$ & $15.993\pm0.034$ & $16.157\pm0.036$ & $16.755\pm0.037$ \\
\noalign{\smallskip}    
$NUV$ & $16.899\pm0.015$ & $16.431\pm0.016$ & $16.555\pm0.031$ & $16.712\pm0.018$ & $17.297\pm0.033$ \\     
\noalign{\smallskip}  
$u$ &
$17.672\pm0.010$ &
$17.053\pm0.007$ &
$17.075\pm0.008$ &
$17.241\pm0.009$ &
$17.837\pm0.012$ \\
\noalign{\smallskip} 
$g$ &
$18.114\pm0.006$ & 
$17.483\pm0.005$ & 
$17.450\pm0.005$ & 
$17.675\pm0.006$ & 
$18.280\pm0.007$ \\
\noalign{\smallskip}  
$r$ &
$18.681\pm0.009$ & 
$18.019\pm0.007$ & 
$17.884\pm0.007$ & 
$18.141\pm0.007$ & 
$18.775\pm0.010$ \\
\noalign{\smallskip}    
$i$ &
$19.075\pm0.014$ &
$18.372\pm0.009$ &
$18.207\pm0.009$ &
$18.518\pm0.010$ &
$19.166\pm0.015$ \\
\noalign{\smallskip}      
$z$ &
$19.413\pm0.051$ & 
$18.747\pm0.034$ & 
$18.505\pm0.035$ & 
$18.860\pm0.034$ & 
$19.477\pm0.058$ \\
\noalign{\smallskip}
$J$  & $$ &   $18.511\pm0.052$ &   $$ & $18.882\pm0.178$ & $19.217\pm0.117$ \\    
\noalign{\smallskip}                                   
$H$  & $$ &   $19.113\pm0.198$ &   $$ & $18.746\pm0.099$  &   $$ \\ 
\noalign{\smallskip}                                   
$W1$ & $$ &   $$ & $$   & $17.305\pm0.130$ & \\ 
\noalign{\smallskip}                    
$W2$ & $$ &   $$ & $$   & $16.282\pm0.192$ & \\ 
\noalign{\smallskip}                    
$W3$ & $$ &   $$ & $$   & $12.128\pm0.269$ & \\ 
\noalign{\smallskip}                    
$W4$ & $$ &   $$ & $$   & $\,\,\,8.250\pm0.190$  & \\ 
\noalign{\smallskip}
\hline
\end{tabular}
\label{tab:mags}
\end{table*}
\twocolumn
}

\section{Discussion}
\label{sect:discussion} 

The high RV variations - partly on short time scales - detected in four of our stars 
are a strong hint that they are part of close binary systems. 
Unlike hot subdwarfs, which are expected to have lost almost their entire envelope
already on the RGB via stable Roche lobe overflow and/or common envelope evolution,
the mass-loss on the RGB of post-AGB close binaries must have been weaker. This is 
because only stars with an sufficiently thick envelope are able to maintain H-shell 
burning and to ascend the AGB.\\
The post-AGB phase (i.e., the evolution from the AGB towards the maximum 
\Teff) is very short-lived. The evolutionary time scales depend strongly on the 
stellar mass, so that a $0.53$\,\Msol\ star needs about 80\,000\,yr to reach its 
maximum \Teff, while a $0.83$\,\Msol\ stars reaches its maximum \Teff\ within less 
than 1000\,yrs \citep{MillerBertolami2015}. It is expected that each post-AGB 
star eventually becomes a CSPN, if the star is able to ionize the previously ejected 
outer layers before the expanding envelope becomes too tenuous to 
appear as a PN.\\
As mentioned above all of our five stars are located in the $\log$ \Teff\ -- \logg\ 
diagram just amongst the CSPNe. However, no nebular lines can be detected in their 
spectra. We will therefore discuss in the following the evolutionary status of these 
objects taking into account the absence of a PN around these stars as well as their 
possible close binary nature.

\subsection{The PG\,1159 star J1556}
\label{subsect:pg1159} 

The phenomenon of the ``missing'' PN is well known for the H-deficient PG\,1159 and O(He) 
stars. Only 15 of the 47 currently known PG\,1159 stars are confirmed CSPNe, which can be 
understood with regard to their evolutionary history. PG\,1159 stars are believed to be 
the result of a (V)LTP that occurs either during the blueward excursion of the post-AGB 
star (LTP), or during its early white dwarf cooling phase (VLTP). The release of nuclear 
energy by the flashing helium shell forces the already very compact star to expand back 
to giant dimensions - the so-called born-again scenario 
\citep{Fujimoto1977, Schoenberner1979, Iben1983, Althaus2005}. The absence of hydrogen in 
the atmosphere of J1556 combined with the appearance of N\,{\sc v} lines in the SDSS 
spectrum suggests a VLTP \citep{wernerherwig2006}. Since for He-burners a three times 
longer evolutionary time scale is predicted \citep{Iben1983} and given that J1556 is going 
through the post-AGB phase for the second time, the ``missing'' PN can be understood very 
well.\\
Should the RV variations in J1556 be confirmed to originate from a close companion, then 
J1556 would be just the second known PG\,1159 close binary system besides 
\object{SDSS\,J212531.92$-$010745.9} \citep{Nageletal2006, Schuh2009, Shimansky2015}. The 
companion in the latter system is a low mass main sequence star as it is obvious from the 
reflection effect in the light curve and from the Hydrogen Balmer series which appears in 
emission due to the radiation of the PG\,1159 star \citep{Nageletal2006}. However, in the 
spectra of J1556, we cannot find evidence for such emission lines. The IR excess of J1556 
showes up only in the WISE bands, which suggest that a possible companion to J1556 might be 
even less massive. We stress that SDSS\,J212531.92$-$010745.9 is at least two orders of 
magnitude less luminous compared to J1556, and thus it is much more difficult to detect a 
low mass main sequence star for the latter.\\
Amongst the H-deficient Wolf Rayet type central stars, which are believed to be the progenitors
of the PG\,1159 stars \citep{wernerherwig2006}, only three close binary systems have been 
discovered so far \citep{Hajduk2010, Manicketal2015}. This leads to a known close binary fraction 
of about 5\% amongst these C-dominated objects\footnote{There are currently about 60 Wolf Rayet 
type central stars known (H.Todt, priv. comm.).}, which is significantly less compared to 
the overall close binary fraction of CSPNe ($\approx 15-20\%$, \citealt{Bond2000, Miszalski2009}).
This suggests, that binary interaction may not play a fundamental role in forming C-dominated stars. 
Definitive statements can, however, only be made after a systematic search of close binaries 
amongst these stars.

\subsection{The O(He) star J0757}
\label{subsect:ohe} 

J0757 is the first RV variable O(He) star ever discovered. \cite{DeMarco2015} recently discovered 
the O(He)-type CSPN of Pa\,5 (\object{PN\,G076.3$+$14.1}) to have a consistent photometric period 
of 1.12\,d in twelve individual Kepler quarters. However, they failed to detect RV shifts larger 
than 5\,km/s. They argue, that the photometric variability may be explained by the influence of an 
orbiting planet or magnetic activity, but most likley by a evolved companion in a nearly-pole-on 
orbit. Given that there are currently only ten O(He) stars known, the confirmation of close binaries 
to J0757 and Pa\,5 would already indicate a binary fraction of 20\%. This may suggest, that 
close binary evolution is, at least for some of the He-dominated objects, an important formation 
channel.\\ 
Also amongst He-dominated stars, a PN seems to be the exception rather than the rule. In addition, it 
appears that the presence of a PN is restricted to N-rich O(He) stars and [WN]-type CSPNe only, which 
suggests that these objects are formed differently than C-rich or C+N-rich He-dominated objects 
\citep{Reindletal2014b}. The possibility that the latter two subclasses are formed via the merger of 
two white dwarfs naturally explains the missing PN, because of the very long predicted post-merger 
times of the stars. Though, the confirmation of a close companion to J0757 would question the merger 
scenario as triple stars systems are arranged hierarchically. Moreover, it is questionable if it is 
possible that unstable mass transfer could stop before the merger of two white dwarfs is complete, 
and thus leaving behind a very low mass close companion \citep{justham2010, Han1999}.\\
We consider therefore yet another possibility to explain both, the ``missing'' PN as well as the 
extreme maximum RV shift of $107.0 \pm 22.0$\,km/s within only 31\,min (which is already pointing 
toward a very close binary system), namely that J0757 could have evolved from an AMCVn type system 
which stopped accretion. AMCVn stars are He transferring ultracompact binary systems with orbital 
period below 1\,h. The mass ratio of these systems is expected to be sufficently low and therefore, 
stable mass transfer occures. These system prevent the merger and with decreasing accretion rates, 
they evolve towards longer orbital periods \citealt{Yungelsonetal2002, Marshetal2004, Nelemans2010}).
The accretores of these sytems may experience stable He-burning \citep{Piersanti2014, Piersanti2015}. 
In case the progenitor of J0757 was a He WD, which ignited He-burning, the post accretion evolution might 
look similar as in the case of the post-double He white dwarf merger models of 
\cite{zhangetal2012b, zhangetal2012a} and, thus, would imply that J0757 must have evolved through the 
region of the central He burning He-sdO stars (lower panel of 
Fig.\,\ref{fig:tefflogg_postagb_Hdef}).\\
Quite interestingly, we find that the great majority (up to 84\%) of the RV variable He-sdO stars 
discovered by \cite{Geieretal2015}, belong to the class of C-rich He-dominated 
objects\footnote{Classification based on the presence of C lines, but absence of N lines in the SDSS 
spectra. The remaining RV variable He-sdO stars either do not show any metal 
lines (8\%) or belong to the group of C$+$N-rich He-sdO stars (8$-$12\%).} just like J0757. Also note 
that the irregular RV variations detected by \cite{Geieretal2015} in some of the He-sdO stars (e.g., 
the C-rich He-sdO \object{SDSS\,J232757.46+483755.2}), might even support this scenario, if we assume 
that they are caused by a still present accretion stream \citep{Schwarz2010}.

\subsection{H-rich stars}
\label{subsect:dis_hrich} 

J1000 and J0935 are the first RV variable naked O(H) stars ever discovered.
So far, only five other naked O(H) stars are known: \object{BD$+18\,2647$} \citep{BauerHusfeld1995}, 
\object{ROB\,162} in the globular cluster NGC\,6397 \citep{Heber1986}, 
\object{vZ1128} in the globular cluster M\,3 \citep{Chayeretal2015}, GD\,605, which is also 
considered as halo star \citep{Fontaine2008}, as well as the metal poor O(H) star \object{LS\,IV-12\,1} 
\citep{HeberHunger1987}.
The fact that no PN is present around these stars is rather unexpected, since in our canonical  
understanding each star that evolves through the AGB is expected to produce a PN. A simple explanation 
for their ``missing'' PNe could be that their post-AGB times are longer than the time it took for the 
PNe to disappear into the interstellar medium.
This possibility is, however, put into doubt, when we have a look in Fig.\,\ref{fig:tefflogg_postagb_H},
where we show in addition to the eight known naked O(H) stars also the locations of H-rich CSPNe found 
in the literature. We derived the post-AGB times from the post-AGB tracks of \cite{MillerBertolami2015} 
for the seven naked O(H) stars as well as the H-rich CSPNe. As it can be seen from Table~\ref{tab:ages}, we 
found that - compared to the eight naked O(H) stars - there are several CSPNe with apparently much 
longer post-AGB times\footnote{Note that for the two DA white dwarf CSPNe \object{HaWe\,5} and 
\object{HDW\,4}, \cite{Napiwotzki1999} even found post-AGB times of several million years, which, 
however, strongly contradicted the very short kinematical ages of both PNe (less than 4\,kyr). 
\cite{Napiwotzki1999} suggested that PNe of these stars might actually be ejected nova shells.}.\\  
On that basis, it is obscure why those stars still harbor a PNe, but not the seven naked O(H) stars and it 
raises the question: Does every post-AGB star, and in particular also every close binary post-AGB 
system, necessarily produce a PNe? This would be particularly important to understand in view of the proposed 
role of close binary post-AGB stars in the formation of asymmetrical PNe. We stress that if J1000 and 
J0935 indeed turn out to be close binary systems, this would lead to a close binary fraction of 25\% 
amongst naked O(H) stars, which is even higher than what is found for the CSPNe.

\begin{table}
\centering
\caption{Post-AGB times as derived from H-burning post AGB tracks of \cite{MillerBertolami2015} of the 
eight naked O(H) stars and some CSPNe with very long post-AGB times. For objects that are not covered by 
these tracks, we give lower limits.}
\begin{tabular}{llr}
\hline\hline
\noalign{\smallskip}
 & name & $t_{\mathrm{post-AGB}}$ \\
\hline
\noalign{\smallskip}
 & LS\,IV-12\,1$^{\mathrm{(a)}}$ & $\,\,\,9$\,kyr \\
 & GD\,605$^{\mathrm{(b)}}$      & $20$\,kyr \\
 & J2046$^{\mathrm{(c)}}$        & $31$\,kyr \\
 & J0935$^{\mathrm{(c)}}$        & $31$\,kyr \\
 & J1000$^{\mathrm{(c)}}$        & $33$\,kyr \\
 & vZ1128$^{\mathrm{(d)}}$       & $38$\,kyr \\
 & ROB\,162$^{\mathrm{(e)}}$     & $49$\,kyr \\
 & BD$+18\,2647^{\mathrm{(f)}}$  & $56$\,kyr \\
\hline
\noalign{\smallskip}
PN\,G & name & $t_{\mathrm{post-AGB}}$ \\
\hline
\noalign{\smallskip}
\object{047.0$+$42.4}  &   A\,39$^{\mathrm{(g)}}$        &  $\,73$\,kyr \\
\object{072.7$-$17.1}  &   A\,74$^{\mathrm{(h)}}$        &  $\,80$\,kyr \\
\object{158.9$+$17.8}  &   PuWe\,1$^{\mathrm{(h)}}$      &  $\,80$\,kyr \\
\object{025.4$-$04.7}  &   IC\,1295$^{\mathrm{(h)}}$     &  $>80$\,kyr \\
\object{030.6$+$06.2}  &   Sh\,2-68$^{\mathrm{(h)}}$     &  $>80$\,kyr \\
\object{158.5$+$00.7}  &   Sh\,2-216$^{\mathrm{(i)}}$    &  $>80$\,kyr \\
\object{077.6$+$14.7}  &   A\,61$^{\mathrm{(g)}}$        &  $\,90$\,kyr \\
\object{128.0$-$04.1}  &   Sh\,2-188$^{\mathrm{(j)}}$    &  $120$\,kyr \\
\noalign{\smallskip}
\object{047.0$+$42.4}  &   A\,39$^{\mathrm{(f)}}$        &  $\,73$\,kyr \\
\object{072.7$-$17.1}  &   A\,74$^{\mathrm{(g)}}$        &  $\,80$\,kyr \\
\object{158.9$+$17.8}  &   PuWe\,1$^{\mathrm{(f)}}$      &  $\,80$\,kyr \\
\object{025.4$-$04.7}  &   IC\,1295$^{\mathrm{(g)}}$     &  $>80$\,kyr \\
\object{030.6$+$06.2}  &   Sh\,2-68$^{\mathrm{(g)}}$     &  $>80$\,kyr \\
\object{158.5$+$00.7}  &   Sh\,2-216$^{\mathrm{(h)}}$    &  $>80$\,kyr \\
\object{077.6$+$14.7}  &   A\,61$^{\mathrm{(f)}}$        &  $\,90$\,kyr \\
\object{128.0$-$04.1}  &   Sh\,2-188$^{\mathrm{(j)}}$    &  $120$\,kyr \\
\noalign{\smallskip}
\hline
\end{tabular}
\tablefoot{Post-AGB times derived from atmospheric parameters, which were taken from 
\tablefoottext{a}{\cite{HeberHunger1987},}
\tablefoottext{b}{\cite{Fontaine2008},}
\tablefoottext{c}{this work,}
\tablefoottext{d}{\cite{Chayeretal2015},}
\tablefoottext{e}{\cite{Heber1986},}
\tablefoottext{f}{\cite{BauerHusfeld1995},}
\tablefoottext{g}{\cite{Napiwotzki1999},}
\tablefoottext{h}{\cite{ZieglerPhD2012},}
\tablefoottext{i}{\cite{Rauch2007},}
\tablefoottext{j}{\cite{Gianninas2010},}
}
\label{tab:ages}
\end{table}

An alternative solution to the non-ejection of a PNe of apparently naked O(H) stars was suggested by 
\cite{HeberHunger1987}, namely that also these stars suffered a LTP. Contrary to the VLTP scenario, which 
always results in a H-deficient surface composition since H is mixed and burnt already during the thermal 
pulse, the LTP does not necessarily predict a H-poor star. 
This is because in the case of a LTP, the convective shell triggered by excessive He burning is not able to 
penetrate the H-rich envelope from below because the entropy jump across the He/H interface is too large. 
Only when the star evolves back to its Hayashi limit on the AGB (\Teff$\lesssim$\,7000\,K), envelope 
convection sets in again \citep{BloeckerSchoenberner1996, BloeckerSchoenberner1997, Schoenberner2008}. 
Evolutionary calculations without overshoot (e.g., \citealt{BloeckerSchoenberner1997}) predict only mild, if any,
mixing. The envelope convection does not reach the layers enriched with carbon, and no third dredge up occurs. 
Later calculations by \cite{Herwig2001} showed that if overshoot is applied to all convective regions, AGB 
models show efficient dredge up even for very low envelope masses and thus produce H-deficient stars. 
Observational proof for the validity of the latter case is offered by \object{FG\,Sge}, the only star known 
to this day that certainly must have suffered a LTP \citep{Schoenberner2008}. This star was actually observed 
evolving back to the AGB were the H fraction in its atmosphere got diluted significantly (from 0.9 to 0.01, by 
number, \citealt{Jeffery2006}).\\
Finally, it is interesting to note that there are also some F or G type supergiants which have already 
departed from the AGB, but lack a reflection nebula, and as such are not expected to evolve into a CSPN.
These much cooler objects (compared to the O(H) stars discussed above) show the presence of a compact 
disk with an inner radius of $\approx 15$\,AU \citep{Deroo2006, Deroo2007} and in addition they were found 
to reside in binary systems. They have orbital periods between a hundred and a couple of thousands days, 
and probably unevolved (very) low initial mass companions \citep{vanWinckel2009}. It is thought, 
that these stars probably did have a pre-PN before, but that the re-accretion of material has stalled the 
blue-ward evolution of the post-AGB star and, thus, providing the circumstellar material enough time to 
disperse before it cloud became ionized \citep{DeMarco2014}. 
 
\section{Conclusions}
\label{sect:conclusions}

Our non-LTE model atmosphere analysis revealed that all five stars, which were recently discovered
by the MUCHFUSS project as RV stars, lie in the post-AGB region of the HRD. 
The RV variations were found not to be significant in case of the O(H) star J2046, but in 
the cases of the other two O(H) stars (J1000, J0935), the O(He) star (J0757), and the 
PG\,1159 star (J1556). This reveals J0757 as the first RV variable O(He) star ever discovered, 
J1556 as the only second known RV variable PG\,1159 star, and J1000 and J0935 as the first RV 
variable O(H) stars that do not show any hint of a PN.\\
The absence of a PN around the PG\,1159 star J1556 can be explained well in terms of a VLTP 
evolution, which can also account for the high C abundance. Given the currently known low binary 
fraction of C-dominated stars, it appears at the moment that binary interactions do not play a 
crucial role in the formation of these stars.\\
In case of the O(He) star J0757 we speculate that it once was part of a compact He transferring binary 
systems in which the mass transfer had stop after a certain time, leaving behind a low mass close 
companion. In this way one could explain both, the absence of a PN as well as the RV shift of 
$107.0 \pm 22.0$\,km/s measured within only 31\,min. Surely, these speculations need more 
observational as well as theoretical support, however, they might offer a yet not considered 
evolutionary channel for the mysterious He-dominated objects.\\
Various explanations could hold for the ``missing'' PNe around the O(H)-type stars. 
The possibility that the post-AGB times of the naked O(H) stars are longer than it took 
for the PNe to disappear into the interstellar medium, seems odd given the fact that there 
are several CSPNe with much longer post-AGB times. Besides the non-ejection of a PN, the 
occurrence of a LTP, or the re-accretion of the PN in the previous post-AGB evolution offer 
possible explanations for those stars not harbouring a PN (anymore).\\
The search for an IR excess was successful only in the case of J1556, which displays a 
clear IR excess in all WISE colors. In order to understand its nature, spectroscopic 
follow-up observations are necessary. We should, however, keep in mind that it is very 
difficult to detect a close companion via the IR excess method given the very high 
luminosities of stars in post-AGB region (our stars have several thousand times the 
solar luminosity, Table~\ref{tab:parameters}). Thus, the presence of a late type 
companion to the other stars in our sample cannot be excluded from the outset.

\begin{acknowledgements}
N.R. was supported by DFG (grant WE\,1312/41-1) and DLR (grant 50\,OR\,1507). 
We thank Marcelo Miguel Miller Bertolami for providing us with the post-AGB tracks before they 
were published. We thank Thomas Rauch and Peter van Hoof for helpful discussions and comments.
Based on observations collected at the Centro
Astronómico Hispano Alemán (CAHA) at Calar Alto, operated jointly by
the Max-Planck Institut für Astronomie and the Instituto de Astrofísica de
Andalucía (CSIC). Based on observations with the William Herschel and Isaac
Newton Telescopes operated by the Isaac Newton Group at the Observatorio
del Roque de los Muchachos of the Instituto de Astrofisica de Canarias on
the island of La Palma, Spain. 
Funding for SDSS-III has been provided by the Alfred P. Sloan Foundation, the 
Participating Institutions, the National Science Foundation, and the U.S. Department 
of Energy Office of Science. The SDSS-III web site is \url{http://www.sdss3.org/}. 
SDSS-III is managed by the Astrophysical Research Consortium for the Participating 
Institutions of the SDSS-III Collaboration including the University of Arizona, 
the Brazilian Participation Group, Brookhaven National Laboratory, Carnegie Mellon University, 
University of Florida, the French Participation Group, the German Participation Group, 
Harvard University, the Instituto de Astrofisica 
de Canarias, the Michigan State/Notre Dame/JINA Participation Group, Johns Hopkins 
University, Lawrence Berkeley National Laboratory, Max Planck Institute for Astrophysics, 
Max Planck Institute for Extraterrestrial Physics, New Mexico State University, New York 
University, Ohio State University, Pennsylvania State University, University of Portsmouth, 
Princeton University, the Spanish Participation Group, University of Tokyo, University of 
Utah, Vanderbilt University, University of Virginia, University of Washington, and Yale 
University. 
This research has made use of the VizieR catalogue access tool, CDS, Strasbourg, France.
\end{acknowledgements}

\bibliographystyle{aa}
\bibliography{muchfuss_postagb} 

\end{document}